\documentstyle[12pt,epsfig]{article}             
\newcommand{\beq}{\begin{equation}}             
\newcommand{\eeq}{\end{equation}}               
\newcommand{\bqry}{\begin{eqnarray}}            
\newcommand{\eqry}{\end{eqnarray}}              
\newcommand{\bqryn}{\begin{eqnarray*}}          
\newcommand{\eqryn}{\end{eqnarray*}}            
\newcommand{\preprint}[1]{\begin{table}[t]      
            \begin{flushright}                  
            \begin{large}{#1}\end{large}        
            \end{flushright}                    
            \end{table}}                        
\newcommand{\PD}[2]                             
    {\frac{\partial^{#2}}{\partial #1^{#2}}}    
\begin{document}
\preprint{LA-UR-99-4171}
\title{Analysis of Dislocation Mechanism \\ for Melting of Elements}
\author{\\ Leonid Burakovsky\thanks{E-mail: BURAKOV@T5.LANL.GOV} 
 \ and \ Dean L. Preston\thanks{E-mail: DEAN@LANL.GOV} \
\\  \\ 
Los Alamos National Laboratory \\ Los Alamos, NM 87545, USA }
\date{ }
\maketitle
\begin{abstract}
The melting of elemental solids is modelled as a dislocation-mediated 
transition on a lattice. Statistical mechanics of linear defects is used to 
obtain a new relation between melting temperature, crystal structure, atomic 
volume, and shear modulus that is accurate to 17\% for at least half of the 
Periodic Table.
\end{abstract}

\bigskip


Phase transitions in many physical systems can be understood in terms of an
abrupt proliferation in the number density of line-like structures. Onsager 
\cite{Ons} suggested in 1949 that the $\lambda $-transition in liquid $^4$He
could be described in terms of the unbinding of vortex lines. The modeling of 
melting as a dislocation-driven phase transition was advocated by Shockley
\cite{Shock}. Edwards and Warner \cite{EW} later constructed a theoretical 
model for dislocation-mediated melting and applied it to the melting of iron. 
More recently, the elementary defect theory of melting of refs.\ [1-3] has 
been developed to a higher degree of sophistication in the textbook by 
Kleinert \cite{KleinII}. Other examples include: (i) the 
superconducting/normal conducting transition in bulk superconductors 
\cite{KleinI}, where the line-like structures are magnetic flux lines, 
(ii) the deconfinement transition in QCD \cite{Patel}, where these structures 
are color flux tubes, (iii) the condensation of synthetic macromolecules 
\cite{Wiegel} and the viral DNA helicity transition \cite{WM}, where the 
line-like structures are macromolecules, and (iv) the glass transition 
\cite{RD}, where linear defects are $Z_2$ disclinations. 

In this paper we will show that linear defects, namely dislocations, drive the 
melting transition and lead to a simple relation, a new dislocation-mediated 
melting law, between melting temperature, crystal structure and elasticity 
data. The relation holds for a surprisingly large number of elements, and 
should also apply to alloys and compounds, such as salts.

Although the energy per unit length, $\sigma ,$ of a dislocation can be very 
large, this energy can always be compensated at sufficiently high temperatures 
by the large entropy of line-like structures. In order to see this, consider 
an ensemble of static non-interacting linear defects with energy per unit 
length $\sigma $ on a $d=3$ dimensional simple cubic lattice of volume $V$ 
and lattice constant $a.$ The grand canonical partition function for such an 
ensemble at temperature $T\equiv 1/\beta $ $(k_B=1)$ is $Z=\exp \;\!\{Z_1\},$ 
where
\beq 
Z_1=\sum _{{\rm strings}}e^{-\beta E}=\sum _Ln(L)e^{-\beta \sigma L}
\eeq 
is the canonical partition function for a single static linear defect, 
and $n(L)$ is the number of configurations of a defect of length $L.$ For 
simplicity, we require the defects to lie along the links of the lattice. 
Then, for $L\gg a$ 
\cite{Copeland}
\beq
n(L)\simeq \frac{V}{a^3}\left( \frac{L}{a}\right) ^{-q-1}\!\!(z')^{L/a},
\eeq
where $z'$ is the number of possible directions that a defect segment can take 
from a given lattice point. $z'=z-1$ if backtracking is not allowed, and 
$z'=z$ otherwise, where $z$ is the coordination number; $z=2d$ for a simple 
cubic lattice in $d$ dimensions. The parameter $q$ accounts for the topology 
of the defect network: $q=-1$ for non-interacting open defects, $q=d/2$ for 
non-interacting closed defects, and $q=7/4$ for non-self-intersecting defects
at low defect densities in 3 dimensions \cite{Copeland}. Thus,
\beq
Z_1\simeq \frac{V}{a^3}\sum _L\left( \frac{L}{a}\right) ^{-q-1}\!\!
e^{-\beta \sigma _{{\rm eff}}L},
\eeq
where 
\beq
\sigma _{{\rm eff}}\equiv \sigma -\frac{T\ln z'}{a}
\eeq
is the effective energy cost to create unit length of a defect at temperature 
$T.$ At the critical temperature $T_{cr}\equiv \sigma a/\ln z',$ there is 
{\it no} energy cost to produce a defect, and therefore defects proliferate 
as $T_{cr}$ is approached from below. At temperatures above $T_{cr},$ the 
divergence of $Z_1$ signals the breakdown of the underlying theory, and the 
system enters a new phase. Hence, the temperature $T_{cr}$ corresponds to a 
phase transition (the particle physics analog of which is the so-called 
Hagedorn transition) in which linear defects are copiously produced. We thus 
consider the melting of elements to be a dislocation-mediated transition, and 
equate the melting temperature, $T_m,$ to $T_{cr}.$ 

Before proceeding further, however, we briefly address the question of the
order of the melting transition in this approach. With a constant dislocation 
energy per unit length, the phase transition discussed above is second order. 
However, in view of Eq.\ (5) below, the dislocation self-energy is expected 
to be a function of dislocation density: it decreases with dislocation density 
because of screening. Although the $\rho $ dependence of the self-energy does 
not follow rigorously from theory, estimates made in \cite{Miz,Cott} show 
that such dependence would result in $-\rho \ln \rho $ dependence of the free 
energy of the dislocation ensemble on dislocation density, and lead to a first 
order melting transition that occurs when the free energy of the crystal 
with a sufficiently high dislocation density equals the free energy of the 
dislocation-free crystal. Recently, this mechanism of a first-order phase 
transition in a system of dislocations has been applied to vortex line 
lattices \cite{KV}.  

An alternative melting mechanism is discussed by Kleinert in his textbook 
\cite{KleinII}, where it is suggested that the interplay of dislocations and 
disclinations is essential for explaining both the strong first-order nature 
of melting and the transition entropy. 
In fact, our framework describes dislocations, which do not exist in liquids. 
Moreover, although a pure dislocation melting can be first order, as discussed
above, it can only explain a transition to a translationally disordered solid,
not a rotationally disordered liquid. Thus, disclinations certainly play a 
role in the melting process. If a dislocation is viewed as a disclination 
dipole \cite{KleinII}, the transition to a translationally disordered 
solid corresponds to the proliferation of the disclination dipoles. The 
translationally disordered solid may then undergo a Kosterlitz-Thouless-type 
vortex-unbinding transition \cite{KT} to a phase of free disclinations, viz., 
a liquid. 
In this paper we do not discuss the full defect model of melting, as a version 
of it is presently available \cite{KleinII}. Instead, we wish to see how close 
we can estimate transition temperatures on the basis of the energy-entropy 
argument of dislocation lines alone, $\sigma _{{\rm eff}}=0$ in Eq.\ (4). We 
hope thereby that the possible shift in the melting temperatures caused by 
the disclinations is small enough to be negligible. 

In order to apply Eq.\ (4) to get the melting temperature, we take the 
dislocation energy per unit length to be that of a dislocation in a complex 
array or tangle of other dislocations. In that case the stress field of a 
given dislocation beyond the mean interdislocation spacing, $R,$ is largely 
cancelled out by the stress fields of the other dislocations in the complex 
array. It then follows that \cite{core}
\beq
\sigma =\kappa \frac{Gb^2}{4\pi }\ln \left( \alpha \frac{R}{b}\right) 
=\kappa \frac{Gb^2}{4\pi }\ln \left( \frac{\alpha }{b\sqrt{\rho }}\right) .
\eeq
Here, $\kappa $ is 1 for a screw dislocation, and is equal to $(1-\nu )^{-1}
\approx 3/2$ for an edge dislocation, $\nu $ being the Poisson ratio. Edge 
dislocations are energetically suppressed. Also, $G$ is the shear modulus, $b$ 
is the average Burgers vector magnitude, and $\alpha $ is a constant of order 
unity. In the second half of this equation we have taken the distance $R$ to 
be approximately equal to $1/\sqrt{\rho },$ where $\rho $ is the dislocation 
density. Eq.\ (5) accounts for both the linear elastic and core energies. 

Using the above expression for the dislocation energy per unit length, the 
formula for the melting temperature of a simple cubic lattice becomes
\beq
T_m=\frac{Gab^2}{4\pi \delta \ln z'},\;\;\;\delta ^{-1}\equiv \kappa \ln 
\left( \frac{\alpha }{b\sqrt{\rho (T_m)}}\right) . 
\eeq

The derivation of the corresponding expressions for the melting temperatures 
of body-centered cubic (bcc), face-centered cubic (fcc), hexagonal 
close-packed (hcp), ..., elements within the framework of statistical 
mechanics of line-like defects would involve greater geometric complexity.
However, we expect that $T_m$ for any lattice is given by the expression (6) 
with $a$ replaced by the cube root of the volume per atom, and a structure 
factor of order one. Also, $z$ retains its meaning as the coordination 
number, that is, the number of nearest-neighbor atoms in a crystal.

We assume that no backtracking is allowed for dislocations, since each 
backtracking would result in a divergence in the linear elastic interaction 
energy between the overlapping segments. We therefore choose $z'=z-1.$ The 
coordination numbers for the elements considered in our analysis are $z=6$ 
for the rhombohedral (rhomb) lattice, $z=8$ for the bcc lattice, and $z=12$ 
for the fcc, hcp, and double hcp (dhcp) lattices. Since the magnitude of the 
Burgers vector is approximately equal to the interatomic distance in all of 
these lattices, $b$ can also be replaced by the cube root of the volume per 
atom and a structure factor of order one. After absorbing factors of order 
one into $\kappa ,$ we obtain our formula for the melting temperature of 
the elements:
\beq
\frac{Gv}{4\pi NT_m\ln (z-1)}=\delta ,
\eeq  
where $v$ is the unit cell volume and $N$ is the number of atoms per unit cell.
The approximations and assumptions leading to this formula are justified
{\it a posteriori} by its accuracy. We will now evaluate the left-hand-side 
of Eq.\ (7) for two-thirds of the elements of the Periodic Table using the 
measured values of $G,$ $T_m$ and lattice constants.

Both $G$ and $v$ in Eq.\ (7) are in general temperature-dependent quantities, 
and so, as with $\rho (T_m),$ their values to be used in (7) should be those
at $T=T_m,$ not the measured values at room temperature. However, measurements
of $G$ and $v$ above 0.8 $T_m$ are available only for Na, K, Al, Cd and Zn 
\cite{SW}. Consequently, we used the values of $G$ and $v$ at room temperature.
The decrease in $G$ with temperature is partially compensated by the increase 
of the volume of the unit cell. Between room temperature and $T_m,$ $G$ drops 
typically by about 20\% \cite{PW}, while the volume increase is of order 5\% 
\cite{LB-a}, so we expect our values of $\delta $ to be $\sim 15$\% high.
 
Elements excluded from our analysis are those that exist in different 
allotropes, e.g., B, Bi, C, H, He, Li, P, S, Si; 
those that have more than 3 polymorphic phases, e.g., Mn, Pu; and those for 
which we could not find the values of one or more of the parameters to be used
in the analysis. For example, we could not find the values of $G$ for At, Br, 
Cl, F, I, N, O, Rn, although the corresponding lattice constants are known. 

For the elements that are included in our analysis, the values of $a$ are 
taken from \cite{LB-a}, except for Fr (estimated using the formulas of 
ref.\ \cite{Fr}). The values of both $T_m$ and $G$ are mostly taken from 
\cite{Gschn}. We also calculate some values of $G$ from those of the single 
crystal elastic constants \cite{LB-c} using known relations \cite{cs}. For 
Pb and Tl, the deviation of the calculated value from the experimental one 
\cite{Gschn} is larger than 50\%, so we use the calculated value based on more 
accurate data \cite{LB-c}. In other cases, the deviation of our calculated 
values from those of \cite{Gschn} is 39\% for Na, and less than 20\% for all 
the remaining materials, so that we use the experimental values. The values of 
$G$ for Ar, Kr, Ne and Xe are also calculated. The values of $T_m$ for Ar, Kr, 
Ne and Xe are from \cite{LB-a}.

The volumes of the unit cells we use are $v=a^3$ for bcc and fcc, $v=a^2c
\sqrt{3}/2$ for hcp and dhcp, and $v=a^3\sin \alpha \sin \beta ,$ $\cos \beta 
=\cos \alpha /\cos (\alpha /2)$ for rhombohedral lattices. 
For polymorphic elements, the conversion temperatures are taken from 
\cite{LB-a}. For all such elements except Sc, lattice constants are measured 
in all solid phases. Eq.\ (7) can be consistently applied to these solid 
phases by extrapolating the values of $G$ to the phase boundaries,
using the empirical relation \cite{PW}
\beq
G(T)\simeq G(0)\left( 1-\gamma \;\!\frac{T}{T_m}\right) ,
\eeq
with $\gamma =0.23\pm 0.08,$ which was calculated by fitting to data on 
thirteen metals. The shear-modulus data on these elements span a sufficiently 
broad range of temperatures, specifically $T=0$ to $T/T_m\stackrel{>}{\sim }
0.4,$ that $\gamma $ can be accurately determined. For polymorphic elements, 
except for Sc, we calculate $\delta $ in the crystal structure at melt (always 
bcc except for Co). Interestingly, $\delta $ generally changes very little 
even when it is evaluated in the lower-temperature solid phases.

Our results for $\delta ,$ i.e., for the ratio  given in Eq.\ (7), are shown 
in Fig.\ 1.

\begin{center}
\vspace{2cm} 
\epsfig{file=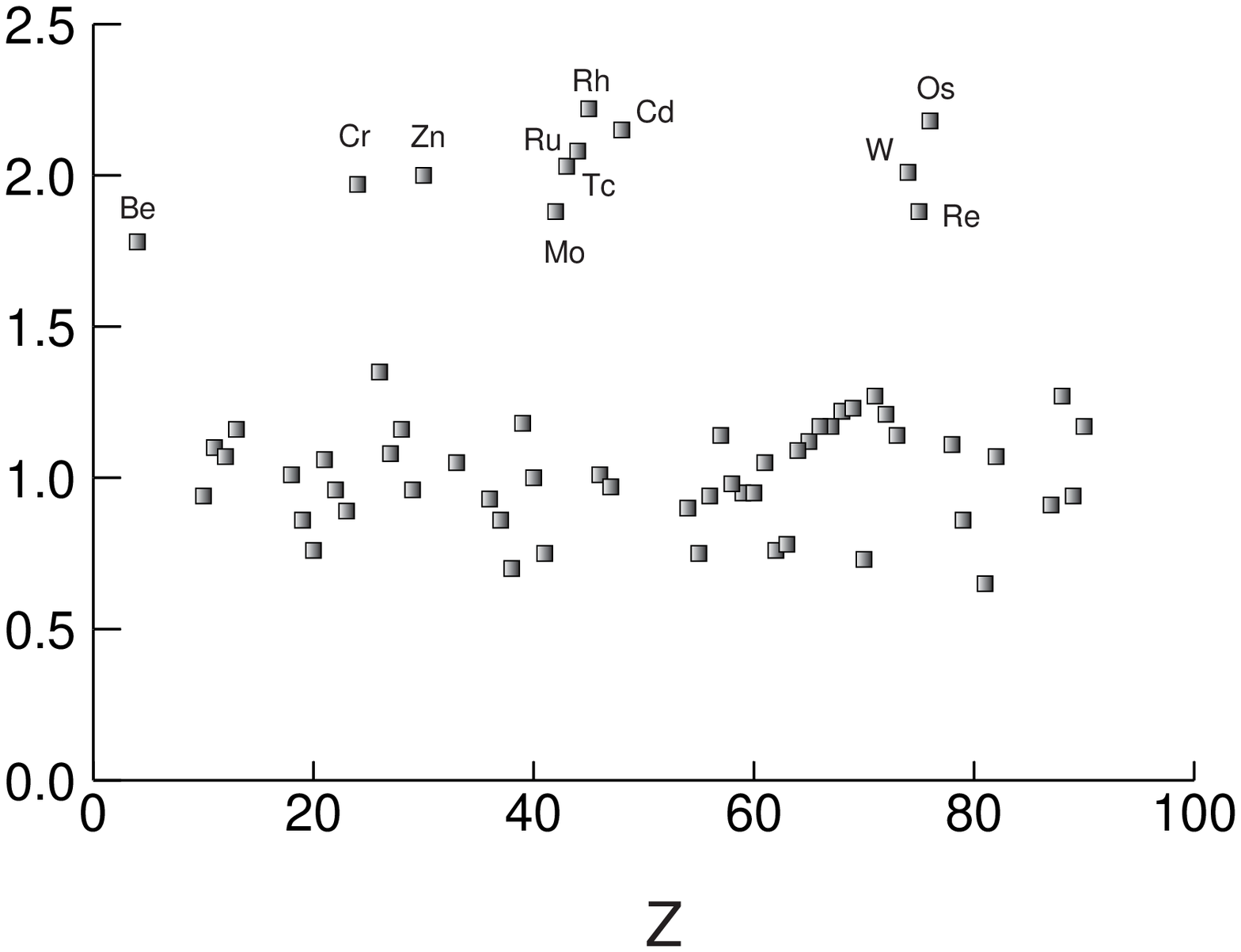,width=15cm,angle=0}
\end{center}
\hspace*{0.3cm} 
Fig.\ 1. Values of $\delta =Gv_{WS}/4\pi T_m\ln (z-1)$ from 
experimental data for 62 elements.  
 \\

The 11 elements that form an outlying group in Fig.\ 1 are explicitly labelled.
We refer to these 11 elements, for which $\delta =2.02\pm 0.13,$ as anomalous, 
and the remaining 51 elements, for all of which $\delta =1.01\pm 0.17,$ as 
regular. We find $\delta =1.19\pm 0.42$ for the entire set of 62 elements.

In Table 1 we make some predictions for the shear moduli of the actinides
Am, Bk, Cm and Es, 
assuming that they all are regular $(\delta =1.0\pm 0.2).$ The melting 
temperatures, and the polymorphic conversion temperatures of Am, Bk and Cm 
(second column) are taken from \cite{actinides}.

\begin{center}
\begin{tabular}{|c|c|c||c|c|c||c|}
\hline 
element & structure & $N$ & $a,$ $\stackrel{\circ }{{\rm A}}$ & $c,$ 
$\stackrel{\circ }{{\rm A}}$ & $T_m,$ K & $G,$ GPa    \\
\hline \hline 
 Am & {\bf dhcp}/1044/fcc/1347/bcc & {\bf 4}/4/2 & 3.468 & 11.240 & 
$1447\pm 2$ & $20.6\pm 4.1$    \\
 Am & dhcp/1044/{\bf fcc}/1347/bcc & 4/{\bf 4}/2 & 4.894 &        & 
$1447\pm 2$ & $21.9\pm 4.4$     \\
 Bk & {\bf dhcp}/1250/fcc & {\bf 4}/4 & 3.416 & 11.069 & 1323 & $19.7\pm 3.9$
   \\
 Bk & dhcp/1250/{\bf fcc} & 4/{\bf 4} & 4.997 &  & 1323 & $22.5\pm 4.5$    \\
 Cm & {\bf dhcp}/1550/fcc & {\bf 4}/4 & 3.500 & 11.134 & 1618 & $22.8\pm 4.6$ 
   \\
 Es & fcc & 4 &  5.746  &   & 1133 & $9.9\pm 2.0$  \\
\hline 
\end{tabular}
\end{center}
Table 1. Predictions for the values of $G$ for the actinides Am, Bk, Cm 
and Es. Boldface indicates the structure used for the calculation of $G$ 
with the help of Eq.\ (7). 
 \\

For the regular 51 elements with $\delta \simeq 1,$ it follows from Eq.\ (6) 
that the dislocation density at melt is
\beq
\rho (T_m)\simeq (e\;\!b)^{-2}\approx (2.7\;\!b)^{-2}.
\eeq
We have set $\alpha =\kappa =1.$ In the case of Cu, $b\approx 2.55\;\!
\stackrel{\circ }{{\rm A}}$ \cite{Cot}, so that $\rho (T_m)\simeq 2\cdot 
10^{14}\;{\rm cm}^{-2},$ which corresponds to a mean separation between 
dislocations at melt of about 7 $\stackrel{\circ }{{\rm A}}.$ For comparison,
consider Cotterill's estimate \cite{Cot} of the threshold dislocation density
above which the atoms in a crystal rearrange into an amorphous state. Under the
assumption that rearrangement does not occur until the amount of slip plane 
between $b$ (approximately equal to the core radius) and $R$ is insufficient 
to withstand the core stress, he derived the inequality $G(R-2b)b\geq 4\pi 
E_{{\rm core}},$ which must be saturated at the threshold dislocation density,
$\rho ^{th}.$ Using the core energy per unit length $E_{{\rm core}}=Gb^2/
4\pi $ \cite{core} (Cotterill used a value proportional to the latent heat of
fusion) we obtain $\rho ^{th}\simeq (3b)^{-2},$ which agrees with our linear 
defect model estimate (9). We have therefore found that dislocation density 
at melt is close to its limiting value as estimated by Cotterill. 

Poirier and Price \cite{PP} identified Zn and Cd, the two anomalous elements 
with relatively low melting temperatures, as irregular, but could find no 
explanation for their behavior, besides their non-ideal hcp structure with 
a large $c/a$ ratio. The irregular properties of Cr, Mo, W, and Be, Cd, Zn 
were also noted in \cite{LB-a}. 
 
The best known melting rule is that of Lindemann, which assumes that
all elements melt when the atomic vibrational amplitude is a fixed fraction, 
$\approx 1/8,$ of the interatomic distance. Wallace \cite{Wal} calculates the 
value of $\ln L$ for 28 elements which he identifies as regular, where $L$ is 
the Lindemann parameter. Based on this value, we calculate the precision of 
the Lindemann rule for these 28 elements to be 12\%. Wallace also calculates 
the value of $\ln L$ for 8 elements which he identifies as irregular (which 
consist of Cr, Mo, W, 3 of our 11 anomalous elements, and 5 additional 
elements which are absent from our analysis of 62 elements). Based on 
Wallace's listing of $\ln L,$ we calculate the precision of the Lindemann 
rule to be 35\% for all 36 elements of ref.\ \cite{Wal}. Our linear defect 
theoretical formula (7) is accurate to 17\% for over at least half of the 
Periodic Table, and to 35\% for the entire set of 62 elements that we 
analyzed. It is therefore not less accurate than the Lindemann rule (taking 
into account a fewer number of regular elements in Wallace's analysis). 

We now offer a possible explanation for anomalous melting. The translationally 
disordered solid phase, which precedes the liquid phase, is represented in 
our model by a scale-invariant distribution of dislocation loops, namely 
$n(L)\propto L^{-q-1}.$ We suspect that for the anomalous metals this unique 
multi-defect state is a poor representation of the solid at the melting point, 
because they actually melt before losing all of their crystalline structure. 
Hence, we expect the anomalous metals to have relatively large (negative) 
liquid correlation entropies, $S_c^{{\rm liq}}.$ Wallace \cite{Wal} has 
identified Cr, Mo and W, all anomalous elements, as having values of 
$S_c^{{\rm liq}}$ outside the normal range. We also expect the melting 
temperatures of the anomalous elements to lie below the values predicted by 
our linear defect model with $\delta =1$ (normal element). This is indeed the 
case -- the melting temperatures of all anomalous elements are suppressed 
$(\delta >1)$ but never elevated $(\delta <1).$ Anomalous melting can be 
accounted for in our model by assuming that it occurs when $\sigma _{{\rm 
eff}}\simeq Gb^2/(8\pi )=O(0.1)$ eV $\stackrel{\circ }{{\rm A}}$$^{-1}.$ 
Of course, other explanations for anomalous melting can be suggested as well. 

To conclude, we have used the methods of statistical mechanics of linear 
defects to derive a new melting rule. It involves only the crystal structure, 
atomic volume, and shear modulus, but despite its simplicity it is accurate 
to 17\% for at least half of the Periodic Table. A preliminary study of our 
melting rule as applied to alloys and compounds indicates an accuracy 
comparable to that for the elements. These results will be presented in a 
longer paper where we will also apply our melting rule to the calculation 
of melting curves.

\section*{Acknowledgements}
We wish to thank R.R. Silbar and T. Goldman for very valuable discussions
during the preparation of this work. One of us (D.L.P.) would like to express
his gratitude to D.C.~Wallace for several very informative discussions on
melting.

%

\end{document}